\documentclass[sigconf, nonacm]{acmart}

\AtBeginDocument{%
  \providecommand\BibTeX{{%
    \normalfont B\kern-0.5em{\scshape i\kern-0.25em b}\kern-0.8em\TeX}}}

\setcopyright{acmcopyright}
\copyrightyear{2021}
\acmYear{2021}

\copyrightyear{2021}
\acmYear{2021}
\setcopyright{acmcopyright}\acmConference[HT '21]{Proceedings of the 32nd ACM Conference on Hypertext and Social Media}{August 30-September 2, 2021}{Virtual Event, Ireland}
\acmBooktitle{Proceedings of the 32nd ACM Conference on Hypertext and Social Media (HT '21), August 30-September 2, 2021, Virtual Event, Ireland}
\acmPrice{15.00}
\acmDOI{10.1145/3465336.3475111}
\acmISBN{978-1-4503-8551-0/21/08}

\usepackage{latexsym}
\usepackage{multirow}
\usepackage{graphicx}
\usepackage{subfig}
\usepackage{soul}
\definecolor{customred}{HTML}{F9D2C6}
\definecolor{customgreen}{HTML}{E1F9C6}

\begin{document}
\fancyhead{}

\makeatletter
\newcommand\footnoteref[1]{\protected@xdef\@thefnmark{\ref{#1}}\@footnotemark}
\makeatother

\title{``A Virus Has No Religion'': Analyzing Islamophobia on Twitter During the COVID-19 Outbreak}

\author{Mohit Chandra}
\authornote{Authors contributed equally to this research.}
\affiliation{%
  \institution{International Institute of Information Technology, Hyderabad}
  \city{Hyderabad}
  \country{India}}
\email{mohit.chandra@research.iiit.ac.in}

\author{Manvith Reddy}
\authornotemark[1]
\affiliation{%
  \institution{International Institute of Information Technology, Hyderabad}
  \city{Hyderabad}
  \country{India}}
\email{manvith.reddy@students.iiit.ac.in}

\author{Shradha Sehgal}
\authornotemark[1]
\affiliation{%
  \institution{International Institute of Information Technology, Hyderabad}
  \city{Hyderabad}
  \country{India}}
\email{shradha.sehgal@students.iiit.ac.in}

\author{Saurabh Gupta}
\affiliation{%
  \institution{Indraprastha Institute of Information Technology Delhi}
  \city{New Delhi}
  \country{India}}
\email{saurabhg@iiitd.ac.in}

\author{Arun Balaji Buduru}
\affiliation{%
  \institution{Indraprastha Institute of Information Technology Delhi}
  \city{New Delhi}
  \country{India}}
\email{arunb@iiitd.ac.in}

\author{Ponnurangam Kumaraguru}
\authornote{Major part of this work was done while Ponnurangam Kumaraguru was a faculty at
IIIT-Delhi}
\affiliation{%
  \institution{International Institute of Information Technology, Hyderabad}
  \city{Hyderabad}
  \country{India}}
\email{pk.guru@iiit.ac.in}
\renewcommand{\shortauthors}{Chandra, et al.}

\begin{abstract}
  The COVID-19 pandemic has disrupted people's lives driving them to act in fear, anxiety, and anger, leading to worldwide racist events in the physical world and online social networks. Though there are works focusing on Sinophobia during the COVID-19 pandemic, less attention has been given to the recent surge in Islamophobia. A large number of positive cases arising out of the religious \textit{Tablighi Jamaat} gathering has driven people towards forming anti-Muslim communities around hashtags like \#coronajihad, \#tablighijamaatvirus on Twitter. In addition to the online spaces, the rise in Islamophobia has also resulted in increased hate crimes in the real world. Hence, an investigation is required to create interventions. To the best of our knowledge, we present the first large-scale quantitative study linking Islamophobia with COVID-19.

  In this paper, we present \textit{CoronaBias} dataset which focuses on anti-Muslim hate spanning four months, with over $410,990$ tweets from $244,229$ unique users. We use this dataset to perform longitudinal analysis. We find the relation between the trend on Twitter with the offline events that happened over time, measure the qualitative changes in the context associated with the \textit{Muslim} community, and perform macro and micro topic analysis to find prevalent topics. We also explore the nature of the content, focusing on the toxicity of the URLs shared within the tweets present in the \textit{CoronaBias} dataset. Apart from the content-based analysis, we focus on user analysis, revealing that the portrayal of religion as a symbol of patriotism played a crucial role in deciding how the Muslim community was perceived during the pandemic. Through these experiments, we reveal the existence of anti-Muslim rhetoric around COVID-19 in the Indian sub-continent.
\end{abstract}




\maketitle

\section{Introduction}

The global Coronavirus (COVID-19) outbreak has impacted people's personal, social, and economic lives. The pandemonium caused due to misinformation, and insufficient preparation to handle such outbreaks has resulted in increased levels of fear, anxiety, and outbursts of hateful emotions among the population \cite{ahorsu2020fear,zandifar2020iranian,montemurro2020emotional}. Consequently, specific communities are being targeted with acts of microaggression, physical and verbal abuse. Social media is flooded with such hateful posts causing online harassment of particular communities \cite{ashby2020initial,ziems2020racism}. Chinese communities and Asians at large have been subjected to racism in the physical world as well as in online spaces due to the COVID-19 origin theories \cite{schild2020go,ziems2020racism,joubin2020anti,coates2020covid}.

Sinophobia has been one of the prime issues during this pandemic, but the surging level of Islamophobia has not received due attention from the research community. During this pandemic, the Islamic missionary movement \textit{Tablighi Jamaat}'s events that happened in multiple countries of South-East Asia have been in the news due to a large number of COVID-19 positive cases being traced back to the attendees of these congregations \cite{news:hrw,news:times, news:nyt}. Unfortunately, this has resulted in a discriminatory movement where the people from the \textit{Muslim} community are being blamed for the virus spread and portrayed as ``human bombs'' and ``corona jihadis'' \cite{news:saj,news:g} leading to trends like \#coronajihad, \#tablighijamatvirus on Twitter.

Islamophobia has been studied broadly in the context of terrorism in the past. Previous works have shown that the Muslim community is often linked to terror and violence \cite{ciftci2012islamophobia}. However, there are other ways in which Islamophobia manifests itself, as has been the case with the ongoing pandemic. This study brings to light the communalism of the COVID-19 pandemic in India. Through multiple quantitative and qualitative experiments, we reveal the existence of anti-Muslim rhetoric around COVID-19 in the Indian sub-continent. In previous studies, it has been shown that real-world crimes can be related to incidents in online spaces~\cite{10.1007/s11280-017-0515-4}. A similar phenomenon has been observed with the hate crimes against the \textit{Muslim} community after the surge COVID-19 cases~\cite{saikia_2020}. In this paper, we propose the \textit{CoronaBias} dataset (data available here~\footnote{\url{https://github.com/mohit3011/Analyzing-Islamophobia-on-Twitter-During-theCOVID-19-Outbreak}}), hoping that it helps take forward the hate speech detection research. Our work has multiple stakeholders, including legislators, social media companies/moderators, and their users.

In this paper, we focus on analyzing the spread of Islamophobia on the social media platform Twitter, especially in the context of the Indian sub-continent. To the best of our knowledge, this is the first quantitative study at the intersection of Islamophobia and COVID-19. Through this paper, we focus on the following research questions:

\begin{itemize}
    \item \textbf{RQ1 :} How did the offline events that happened during the course of this study statistically affect online behaviour? How did the context associated with the \textit{Muslim} community change over the period of our study?

    \item \textbf{RQ2 :} Which topics were prevalent in the tweets concerned with the \textit{Muslim} community. Were some topics more prevalent than others during a particular window of time?
    
    \item \textbf{RQ3 :} What were the differentiating characteristics of users who were indulged in spreading Islamophobia in contrast to the other users?
    
    \item \textbf{RQ4 :} What was the role played by external sources, especially the news media outlets? What was the nature of the content that was referenced in the tweets through URLs?
\end{itemize}

\noindent\textbf{Our contributions:} In this paper, we present the first of its kind ``\textit{CoronaBias}'' dataset with over $410,990$ tweets from $244,229$ unique users, which we will publicly release on the acceptance of this paper. Additionally, we present a long-term longitudinal analysis. Our key contributions are summarised as follows: i) Drawing a correlation between real-world events and corresponding changes on Twitter in a statistically sound manner, with the help of Temporal Analysis using the PELT algorithm. We also find a growing association of the \textit{Muslim} community with the COVID-19 pandemic through the semantic similarity experiment; ii) Extracting the prevalent topics of discourse in broad and focused windows of time through \textit{Macro} and \textit{Micro} topic modelling experiment. Qualitatively, we find a blend of mixed sentiments over the topics; iii) We surface the differentiating characteristics of users sharing Islamophobic content from other users. We specifically analyze user network, user activity, and descriptions of user bios to reveal the differentiating characteristics; iv) Comparing the nature of external content that is referenced on Twitter via URLs. In addition to presenting the content referenced through URLs from YouTube, BBC, and OpIndia (top shared domains), we present a comparative study of the toxicity of the referenced content.


\section{Related Work}

The unprecedented nature of the ongoing pandemic has attracted much attention from the research community. This section presents some of the past works on research related to COVID-19 and hate speech.

\noindent\textbf{COVID-19 and Online Social Media: } The presence of a massive amount of data related to COVID-19 on various social media platforms has attracted researchers to collect it for different studies \cite{info:doi/10.2196/19273, vidgen2020detectingasian, schild2020go}. \citet{info:doi/10.2196/19273} proposed a multilingual coronavirus Twitter dataset with over 123 million tweets. \citet{schild2020go} presented one of the first studies related to the Sinophobic behavior during the COVID-19 pandemic. The researchers collected over 222 million tweets and performed experiments revealing the rise in Sinophobia. Works done in the past were related to the characterization of online discussion on different issues. Previous works like \citet{ziems2020racism, schild2020go} have focused on a quantitative longitudinal analysis of Sinophobia that emerged during this pandemic. On a different front, \citet{kouzy2020coronavirus} have presented a study to analyze the magnitude of misinformation related to COVID-19 on Twitter. Along similar lines, \citet{ferrara2020covid} have emphasized the role of bots in spreading conspiracy theories related to the pandemic on Twitter. The topic of privacy \& policy in the health sector has been a central focus during this pandemic, primarily due to the importance of contact tracing~\cite{cho2020contact,halpern2020cognitive}. Although there have been quite a few works exploring different aspects of the ongoing pandemic, to the best of our knowledge there is no quantitative analysis based work exploring Islamophobia during the COVID-19 pandemic. We fill this gap in this paper.

\noindent\textbf{Research related to hate speech and Islamophobia: } Hate speech has been a popular topic in the research community since a long time. While there have been works that have focused on the general hate speech detection \cite{davidson2017automated, founta2019unified,elsherief2018hate,chandra-etal-2020-abuseanalyzer} there are others that have focused on specific forms of hate speech like racism \&sexism \cite{waseem2016hateful} and cyberbullying \cite{chatzakou2017measuring}. In another line of work, researchers have used deep learning for hate speech detection \cite{badjatiya2017deep,park2017one}.

Unlike other forms of hate speech, Islamophobia has not been explored in depth. \citet{soral2020media} presented that social media users were exposed to a higher level of islamophobic content than the users who consumed news from traditional forms of media. In another work, \citet{vidgen2020detecting} proposed an automated software tool that distinguishes between non-Islamophobic, weak Islamophobic, and strong Islamophobic content.

\section{Ethical Concerns}
\label{sec:ethical_concerns}
We have used two sources of data in this work: i) COVID-HATE from Twitter (~\citet{info:doi/10.2196/19273}), and ii) News articles scraped from BBC \& OpIndia, and video titles and descriptions from YouTube. The data from Twitter was collected from a publicly available source and we hydrated the tweet IDs using the official Twitter API. Information about YouTube videos was collected with the official YouTube Data API \footnote{\url{https://developers.google.com/youtube/v3}}. For the news articles, we provided a user agent string that made our intentions clear and provided a way for the administrators to contact us with questions or concerns. We requested data at a reasonable rate and strived never to be confused for a DDoS attack. We saved only the data we needed from the page. The use of data in this research warranted a justification of the ethical implications and the decisions made during the work. The methods followed during the study are influenced by the works around the internet research ethics and, more specifically, Twitter research ethics \cite{zimmer2017internet,vitak2016beyond,fiesler2018participant}. We handled data and made decisions in data collection in an ethical manner with no intent to harm any individual's privacy in any way \cite{townsend2016social,mislove2018practitioner,markham2012ethical}. 

\section{\textit{CoronaBias}: An anti-Muslim hate dataset}
\label{section:data}
For our study, we used the data collected from Twitter, ~\citet{info:doi/10.2196/19273} have presented a large corpus using specific keywords related to COVID-19 and have published the tweet IDs on an online repository.\footnote{\url{https://github.com/echen102/COVID-19-TweetIDs}}. We filtered these only to consider tweets written in Roman script (English) from February 1, 2020, till May 31, 2020. Since our work focuses on the \textit{Muslim} community, we further filtered out tweets based on a curated set of keywords. We used positive, negative, and neutral terms so as to avoid bias in our dataset. Some examples of keywords of each type include \textit{tablighiheroes, muslimsaviours} for positive keywords, \textit{muslimvirus, coronajihad} as negative keywords and \textit{islam, muslim} for neutral keywords. (Link to the entire keyword list can be found here~\footnote{\url{https://github.com/mohit3011/Analyzing-Islamophobia-on-Twitter-During-theCOVID-19-Outbreak/blob/main/keywords-major.txt}}).

The filtered dataset (referred as ``\textit{CoronaBias} dataset'' from now on) contains $410,990$ tweets with a month-wise distribution of number of tweets being: February: $105,974$,  March: $70,682$,  April: $161,780$, May: $72,554$. $244,229$ unique users in the dataset have tweeted at least once during the course of the study, out of which $2,107$ are verified accounts. The average word length for the tweets in the dataset is 34. 

To answer \textbf{RQ4}, we collected data from 764 videos on YouTube, 231 unique articles on OpIndia, and 82 unique articles on BBC, which were referenced as URLs in the tweets present in our dataset (ethical concerns related to the data collection have been discussed in Section~\ref{sec:ethical_concerns}).

\subsection{Dataset Annotation}

In addition to creating the \textit{CoronaBias} dataset, we assigned binary labels to the tweets based on the stance of Islamophobia. In this labelling task, we assigned each tweet one of the two labels - `Hateful' or `Non-Hateful'. The sheer number of tweets made it impossible to annotate the entire dataset manually; therefore, we approached the problem of semi-automatic annotation of tweets in three ways -- 1) Using Linguistic Inquiry and Word Count (LIWC) ~\cite{tausczik2010psychological} 2) Using Bidirectional Encoder Representations from Transformers (BERT) based method, 3) SVM based method.

\begin{table}[h]
    \centering
    \small
    \begin{tabular}{|p{0.15\columnwidth}|p{0.15\columnwidth}|p{0.15\columnwidth}|p{0.15\columnwidth}|p{0.15\columnwidth}|}
        \hline
        \textbf{Method} & \textbf{Accuracy} & \textbf{Recall} & \textbf{Precision} & \textbf{F-1}  \\
        \hline
        \hline
        BERT & $\mathbf{0.854 \pm }$ $\mathbf{0.088}$ & $\mathbf{0.854 \pm }$ $ \mathbf{ 0.087}$ & $\mathbf{0.854 \pm }$ $ \mathbf{0.086}$ & $\mathbf{0.853 \pm}$ $\mathbf{0.088}$ \\
        \hline
        SVM & $\mathbf{0.799 \pm }$ $\mathbf{0.024}$ & $\mathbf{0.803 \pm }$ $ \mathbf{ 0.024}$ & $\mathbf{0.806 \pm }$ $ \mathbf{0.023}$ & $\mathbf{0.799 \pm}$ $\mathbf{0.024}$ \\
        \hline
    \end{tabular}
    \caption{Results for the task of tweet classification into `Hateful' and `Non-Hateful' category using BERT and SVM (Using 5-fold Cross-Validation). We report the macro-averaged values for each metric.}
    \label{tab:bert_model_results}
\end{table}

Out of the three approaches, LIWC based method performed the worst. Alternatively, we used SVM and a BERT based model for this task due to its recent success in various downstream NLP tasks \cite{devlin2018bert,xu-etal-2020-discourse,xu2019bert,reimers2019sentencebert}. We used a 2-layer MLP with dropout following the BERT module to classify the data, keeping the learning rate at $10^{-6}$ and dropout value at $0.2$. We fine-tuned the BERT based model on our dataset for maximum of 20 epochs (code~\footnote{\url{https://github.com/mohit3011/Analyzing-Islamophobia-on-Twitter-During-theCOVID-19-Outbreak/tree/main/bert_files}}). We also experimented with SVM where we used a linear kernel along with regularization parameter$=0.1$ and maximum iteration$=3500$.

To create the dataset used for training our models, we filtered out 2000 tweets from the \textit{CoronaBias} dataset using a curated list of keywords that contained positive, negative, and neutral \textit{Muslim}-related terms, to ensure a balanced training data. We used a modified version of the guidelines provided in~\cite{vidgen2020detecting} for the annotation process. The filtered tweets were annotated on binary labels -- 1) Non-Hateful and 2) Hateful according to the stance of tweets concerning the \textit{Muslim} community, by three annotators. The Fleiss' Kappa score for the annotation was $0.812$, which translates to a near-perfect agreement among the annotators.

After the annotation process, the distribution of the number of tweets came out as: Hateful = $1,051$, Non-Hateful = $949$. We used this labelled dataset to train the BERT based model and SVM classifier. Table~\ref{tab:bert_model_results} presents the results for the classifiers. We chose BERT based approach due to its better performance. After training our BERT-based classifier, we classified all the tweets present in the \textit{CoronaBias} dataset as `Hateful' or `Non-Hateful'. We use the classified tweets in the \textit{CoronaBias} dataset for further experiments.

\section{Temporal Analysis}

To answer the first part of \textbf{RQ1} concerning the relation of offline and online events, we carried out the temporal analysis of the annotated tweets present in the \textit{CoronaBias} dataset. In particular, we focused on the correlation between real-world events and the behaviour on Twitter during the course of this study along with the difference in the dynamics between `Hateful' and `Non-Hateful' labelled tweets. We used a methodology similar to the one proposed by \citet{Zannettou_Finkelstein_Bradlyn_Blackburn_2020}. We used change-point analysis to rank each change in the time-series graph based on mean and variance for both the tweet categories. For the change-point analysis, we used the PELT algorithm as described in \cite{killick2012optimal}.

\begin{table*}[ht]
    \centering
    \small
    \begin{tabular}{|p{0.20\columnwidth}|p{0.20\columnwidth}|p{1.5\columnwidth}|}
    \hline
    Change Point Event Rank & Date (MM-DD-YYYY) & Event Description\\
    \hline
    \hline
    1 & 03-31-2020 \& 04-01-2020 & On 31st March, reports linked the \textit{Tablighi Jamaat} congregation event that happened in Delhi to the sudden spread of COVID-19 across the country. \\
    \hline
    1 & 04-26-2020 & COVID Explosion in Maharashtra and record new cases. 24th April marked the start of Ramadan. \\
    \hline
    2 & 04-12-2020 & On 14 April, India extended the nationwide lockdown till 3 May. On 18 April, the Health ministry announced that 4,291 cases were directly linked to the \textit{Tablighi} event. \\
    \hline
    3 & 04-06-2020 & As of 4 April, about 22,000 people who came in contact with the \textit{Tablighi Jamaat} missionaries had to be quarantined. \\
    \hline
    3 & 03-17-2020 & Malaysia reported its first two deaths. By 17 March, the Sri Petaling event had resulted in the biggest increase in COVID-19 cases in Malaysia, with almost two thirds of the 673 confirmed cases in Malaysia linked to this event. \\
    \hline
    4 & 02-26-2020 & On the 25th of February, the Iranian government first told citizens that the U.S. had "hyped COVID-19 to suppress turnout" during elections, and that it would "punish anyone" spreading rumors about a serious epidemic. \\
    \hline
    4 & 05-26-2020 & 24th May was end of Ramadan. \\
    \hline
    \end{tabular}
    \caption{Table linking the ranked change points to the set of offline/real-world events. The ranked change-points are computed using the penalty factor in the PELT algorithm (i.e, higher the penalty, higher is the significance of the change-point and lower is the event rank).}
    \label{tab:event_table}
\end{table*}

PELT is an exact algorithm for change-point analysis that determines the points in time at which the mean and variance changes by maximizing the likelihood of the distribution given the data. We ran the PELT algorithm on the tweets labelled as `Hateful'; additionally, we used a penalty with the algorithm to keep the count of change-points limited. We ran the algorithm with different values of penalty constant ($1-10$) in decreasing order and kept track of the largest penalty amplitude at which each change-point first appeared. This provided us a ranking of the change-points in order of their significance (lower the rank, higher is the significance of the event). For correlating the change points to the real-world events, we considered a window of 10 days around the date of change-point and collected the information of the concerned events through multiple media outlets and Wikipedia. Table~\ref{tab:event_table} comprises event descriptions along with the change-point significance rank and the date of event occurrence.

\begin{figure*}[!htb]
\centering
\minipage{0.80\columnwidth}
\centering
  \includegraphics[width=1.1\columnwidth]{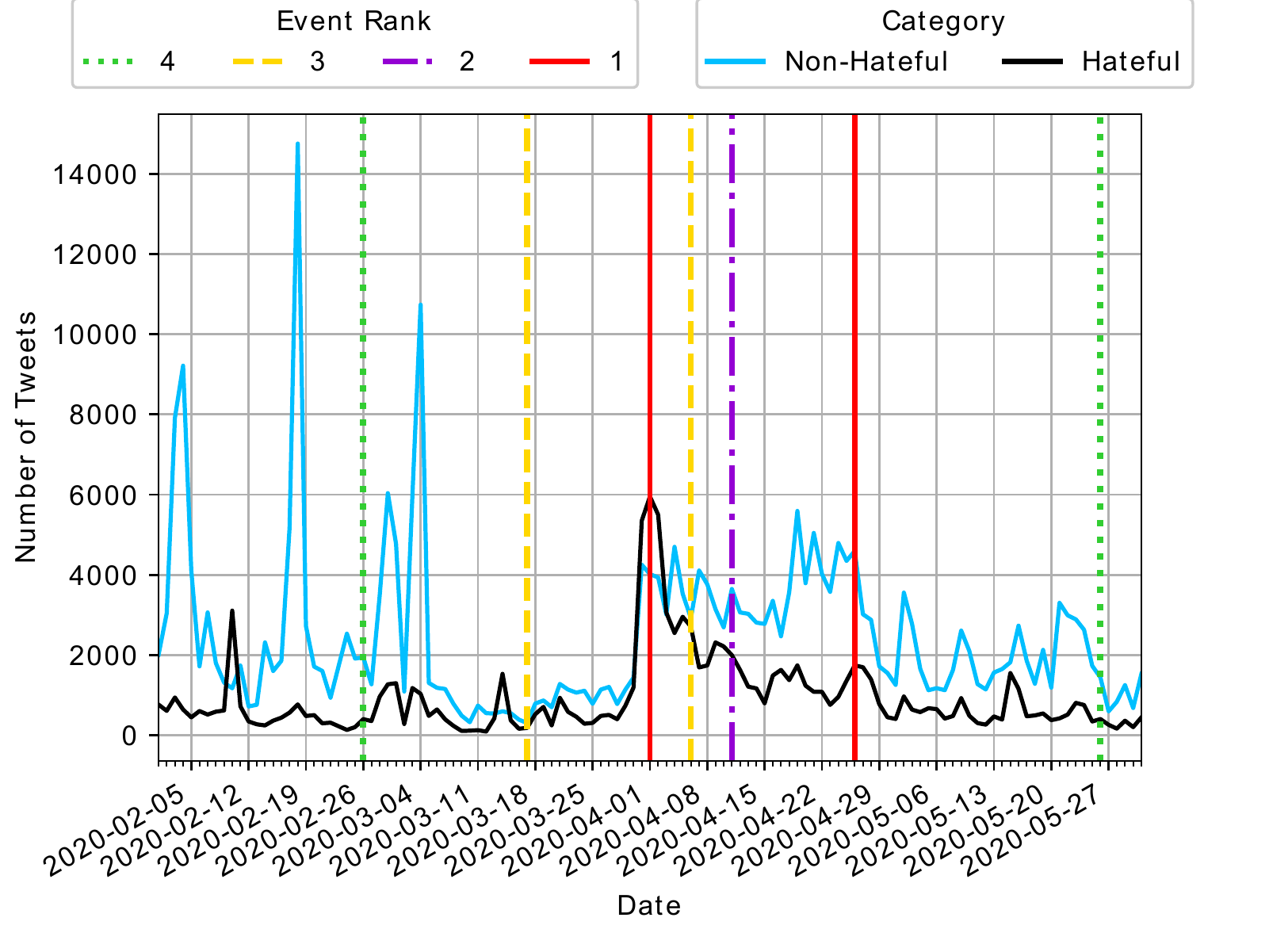}
  \caption{Distribution of `Hateful' and `Non-Hateful' tweets made by the users. The vertical lines represent the event ranks referenced in Table ~\ref{tab:event_table}.}
  \label{fig:temporal_analysis_graph}
 \endminipage\hfill
 \minipage{1.2\columnwidth}
    \captionsetup{type=table}
    \centering
    \small
    \begin{tabular}{|p{0.09\columnwidth}|p{0.09\columnwidth}|p{0.09\columnwidth} |p{0.09\columnwidth}||p{0.09\columnwidth}|p{0.09\columnwidth}|p{0.09\columnwidth}|p{0.09\columnwidth} |p{0.09\columnwidth}|p{0.09\columnwidth}|}
    \hline
    \multicolumn{4}{|c||}{\textbf{Muslim}} & \multicolumn{4}{c|}{\textbf{Virus}} \\
    \hline
    February & March & April & May & February & March & April & May \\
    \hline
    \vtop{\hbox{\strut Syste-}\hbox{\strut matic}} $(0.318)$ & Islam $(0.434)$ & Islam $(0.444)$ & People $(0.457)$ & Disease $(0.447)$ & \vtop{\hbox{\strut Corona-}\hbox{\strut virus}} $(0.446)$ & \vtop{\hbox{\strut Corona-}\hbox{\strut virus}} $(0.491)$ & \vtop{\hbox{\strut Corona-}\hbox{\strut virus}} $(0.540)$ \\
    \hline
    Travelers $(0.305)$ & Bully $(0.288)$ & People $(0.389)$ & Islam $(0.432)$ & Desease $(0.415)$ & Disease $(0.411)$ & Covid19 $(0.427)$ & Covid19 $(0.481)$ \\
    \hline
    Bully $(0.294)$ & \cellcolor{customred}China $(0.283)$ & \cellcolor{customred}China $(0.346)$ & \cellcolor{customred}China $(0.402)$ & Creator $(0.361)$ & Desease $(0.380)$ & \cellcolor{customred}Muslims $(0.406)$ & Disease $(0.471)$ \\
    \hline
    Dungan $(0.282)$ & Moslem $(0.281)$ & \cellcolor{customred}Virus $(0.329)$ & \cellcolor{customred}Covid19 $(0.386)$ & Wrath $(0.350)$ & \cellcolor{customred}Muslims $(0.369)$ & Corona $(0.382)$ & \cellcolor{customred}Muslims $(0.457)$ \\
    \hline
    Reality $(0.280)$ & \cellcolor{customred}Virus $(0.275)$ & \cellcolor{customred}Covid19 $(0.314)$ & \cellcolor{customred}\vtop{\hbox{\strut corona-}\hbox{\strut virus}} $(0.361)$ & Coughing $(0.343)$ & \cellcolor{customred}Allah $(0.342)$ & Disease $(0.363)$ & Corona $(0.434)$ \\
    \hline
    Moslem $(0.273)$ & Seeing $(0.233)$ & US $(0.293)$ & \cellcolor{customred}India $(0.344)$ & Diseases $(0.332)$ & \cellcolor{customred}China $(0.2817)$ & People $(0.349)$ & \cellcolor{customred}Muslim $(0.379)$ \\
    \hline
    \vtop{\hbox{\strut Indige-}\hbox{\strut nous}} $(0.268)$ & People $(0.224)$ & Americas $(0.292)$ & One $(0.329)$ & Source $(0.296)$ & \cellcolor{customred}Muslim $(0.275)$ & \cellcolor{customred}India $(0.334)$ & People $(0.379)$ \\
    \hline
    \end{tabular}
    \caption{Most similar words to \textit{Muslim} and \textit{Virus} in decreasing order of the value of cosine similarity value (in parenthesis). We observed an increase in cosine similarity score between words like \textit{Muslim}, \textit{Virus} and \textit{China} indicating that the context around the pandemic moved towards Islamophobia with the due course of time. }
    \label{tab:word_similarity}
    \endminipage
\end{figure*}

Figure~\ref{fig:temporal_analysis_graph} presents the frequency of tweets present in the `Hateful' and `Non-Hateful' categories from February till May 2020. As expected, the number of 'Non-Hateful' tweets was higher than the number of 'Hateful' tweets in general because the prior category contains both \textit{pro-Muslim} and \textit{Muslim-neutral} tweets.

As observed in Table~\ref{tab:event_table} and Figure~\ref{fig:temporal_analysis_graph}, several change-points coincide with the increased number of `Hateful' tweets on Twitter. Table~\ref{tab:event_table} includes the significant offline events that took place in the real-world around the time of each of the online change points of our dataset.
After looking at two of the highest significant events (event rank 1) in Table~\ref{tab:event_table}, we observed a local maxima in the number of `Hateful' labelled tweets (in fact the number attained a global maxima on 31st March / 1st April 2020). Interestingly, the rank one event on 31st March/1st April is one of the few occurrences in the entire study where the number of `Hateful' tweets was more than the number of `Non-Hateful' tweets. In contrast to that, another event rank 1 (26-04-2020) in Table~\ref{tab:event_table} before which we observed a drop in the number of `Hateful' tweets marked the start of \textit{Ramadan}.

We observed a clear correlation between the offline events and the corresponding behavioural changes on Twitter through the temporal analysis. Additionally, using the PELT algorithm, we surfaced the changes in the statistical properties (mean \& variance) of the trend on Twitter following the real world in a ranked fashion. We observed that offline events with greater significance impacted the online trends more.

\section{Semantic Similarity Experiment}

In the second part of \textbf{RQ1}, we investigated the change in context associated with the \textit{Muslim} community during the pandemic. For this, we performed the semantic similarity experiment. Through this experiment we also wanted to validate our hypothesis that with the due course of time, the \textit{Muslim} community was blamed for the mishappenings during the pandemic. We modified the methodology presented in ~\cite{schild2020go} and trained Word2Vec Continuous-Bag-of-Words (CBOW) model \cite{mikolov2013efficient}.

We chose two general words, `\textit{Muslim}' and `\textit{Virus}' as these can be used in many contexts (referred to as reference words in the rest of this section). We trained a separate Word2Vec model for each month from February till May 2020; more formally, we defined the time period of the study as $\mathcal{T} = \{Feb, March, April, May \}$. For each of the Word2Vec model $\mathcal{W}_{i}, i \in \mathcal{T} $, we computed the most similar words to the two reference words (`\textit{Muslim}' and `\textit{Virus}') using the cosine similarity criterion. Table ~\ref{tab:word_similarity} presents the most similar words for each of the reference words across all the months. Observing the trend for the word \textit{Muslim}, we saw that the list of similar words in February was not related to COVID-19, with a few words being racist slurs. However, we observed the appearance of words like `China', `Virus', `COVID19' in the subsequent months, suggesting that the context around the word `\textit{Muslim}' was used in a context similar to the aforementioned words in the dataset. Similar was the case with the reference word `\textit{Virus}' where we could observe the word being related to various disease-related terminologies and symptoms in the initial two months. However, in the month of March and onwards, we observed the words `Allah' and `Muslims', indicating that people started associating COVID-19 with the \textit{Muslim} community.

Additionally, in Table~\ref{tab:word_similarity} we observed that for the word `\textit{Muslim}', the cosine similarity score of words like `China', `Virus' increased from March to May. Similar was the case with the word `\textit{Virus}' where we observed the cosine similarity score for `Muslims' steadily rise for the same time period, indicating the context around the pandemic inclined more towards Islamophobia due course of time. We also noticed the association of the word `Muslim' with less familiar words like `Dungan', `Indigenous', and `Reality'. A closer look at tweets containing these words revealed the reason behind this association. The word `Dungan' is a term used in the former Soviet Union territories to refer to a group of Muslim people. The word `Reality' was used in many tweets to describe the reality of Islamophobia on the ground. The word `Indigenous' was often used to portray the association of the \textit{Muslim} community being indigenous to India but still facing religious discrimination.

Through this experiment, we showed that the context around the \textit{Muslim} community changed through the course of the pandemic. As time progressed, we observed growing association between the words like `Virus', `COVID19' with the reference word \textit{Muslim} which is indicative of a rising Islamophobic sentiment.

\section{Topic Modelling}

To answer \textbf{RQ}$\mathbf{2}$, we delve deep into the content analysis of the collected data through topic modelling. We divided this experiment into two levels of granularity -- 1) \textit{Macro Level Topic Modelling}, 2) \textit{Micro Level Topic Modelling}. At the macro level, we performed the task of topic modelling on the \textit{CoronaBias} dataset to reveal the prevalent topics across the period of our study. At the micro-level, we chose two significant events during our study and performed the task of topic modelling over a window of ten days around these events.

We experimented with three different approaches -- 1) Latent Dirichlet Allocation (LDA), 2) Top2Vec \citet{angelov2020top2vec}, 3) Non-Negative Matrix Factorization (NMF), out of which NMF based method provided the best results. Due to the limited space, we only discuss the results obtained using the NMF based technique.

\begin{table*}[ht]%
\centering
\small
\begin{tabular}{|l|l|p{0.25\textwidth}|p{0.48\textwidth}|p{0.06\textwidth}|}
\hline
& S.No. & \multicolumn{1}{|c|}{\textbf{Topic}}& \textbf{Tokens} & \textbf{No. of Tweets}\\
\hline
\hline
\multirow{4}{*}{\rotatebox{90}{\textit{Non-Hateful}\ \ \ \ \ \ \ \ \ \ }}&1&\cellcolor{customgreen} The Coronavirus pandemic is distracting people from the Uyghur \textit{Muslim} crisis. & \cellcolor{customgreen} coronavirus distraction, muslims tortured, imprisoned raped, concentration killed, thousands chinese & $5,368$ \\
\cline{2-5}
&2& \cellcolor{customgreen} The Coronavirus Pandemic leading to a global market collapse. & \cellcolor{customgreen} markets collapse, global demand, corona hits, find difficult, attention needed & $1,975$\\
\cline{2-5}
&3& \cellcolor{customgreen} General measures to prevent Covid & \cellcolor{customgreen} washing hands, covering face, five times, shaking hands, covering & $4,470$\\
\cline{2-5}
&4& \cellcolor{customgreen} \textit{Tablighi Jamaat} members donating plasma & \cellcolor{customgreen} donate plasma, tablighijamaat members, members recovered, covid donate, patients & $7,555$\\
\hline
\hline
\multirow{4}{*}{\rotatebox{90}{\textit{Hateful}\ \ \ \ \ \ \ \ \ \ \ \ \ \ \ }}&5 & \cellcolor{customred} Anti-Abrahamic religious sentiments attached with COVID-19. & \cellcolor{customred} kill muslims, china kill, hate muslims, hate christians, hate jews & $6,089$\\
\cline{2-5}
&6 &\cellcolor{customred} \textit{Muslims} attacking Covid-19 health workers & \cellcolor{customred} indian muslims, fight, pandemic, doctors, poor, violence, govt & $28,747$\\
\cline{2-5}
&7 & \cellcolor{customred} \textit{Tablighi Jamaat} religious congregation & \cellcolor{customred} tablighi jamaat, maulana, jamaat attendees, police, jamaat members, pakistan, markaz  & $24,718$\\
\cline{2-5}
&8 & \cellcolor{customred} Assigning blame to \textit{Muslims} and Allah for the spread of Covid-19 & \cellcolor{customred} corona jihad, spreading corona, allah, muslims, corona virus, people  & $19,561$\\
\hline
\end{tabular}
\caption{ Some prevalent topics in \textit{Macro} topic modeling experiment for the entire corpus. The rows in green (1-4) and red (5-8) represent the non-hateful and hateful topics respectively.}
\label{tab:macro_topic_modelling}
\end{table*}

For the NMF based method, we pre-processed the tweets (removal of punctuation, non-ASCII characters, stopwords, and whitespaces, followed by lowercasing). To prevent any anomaly in the method, we removed all the topics that occurred in more than $\sim 85\%$ or less than $\sim 3\%$ of the tweets. For finding the number of topics for both \textit{Macro level} and \textit{Micro level} topic modelling, which could best represent the data, we used the \textit{Coherence score}. We iterated through the number of topics from 5 to 50 with a step size of 5. We then used TF-IDF based vectorization to create tweet vectors which were used in the NMF based method (code here~\footnote{\url{https://github.com/mohit3011/Analyzing-Islamophobia-on-Twitter-During-theCOVID-19-Outbreak/tree/main/topic_modelling_files}}). Furthermore, we classified the topics as \textit{Hateful} and \textit{Non-Hateful} on the stance of portrayal of \textit{Muslims} and the Islamic religion related to COVID-19.

Table~\ref{tab:macro_topic_modelling} presents the prevalent topics along with the tokens and the number of tweets belonging to the particular topic for the \textit{Macro} topic modelling.  Among the topics which were categorized as `Non-Hateful', we observed a high number of tweets for topic 1, which talks about the COVID-19 pandemic distracting people from the issue of torture on Uyghur \textit{Muslims}. Topic 2 talks about the financial crisis and market collapse during the pandemic,  while Topic 3 represents tweets related to general precautions \& measures against COVID-19. Topic 4, interestingly, talks about members of \textit{Tablighi Jamaat} donating plasma for plasma therapy and helping others, tweets belonging to this topic were very prevalent during the month of \textit{Ramadan}. 

In contrast, when we looked at the `Hateful' topics, we found a lot of them related to the different aspects of the \textit{Tablighi Jamaat} event and its members. Topic 5 represents anti-Abrahamic religious sentiments that arose due to various conspiracy theories related to the origin of COVID-19. Topic 6 is a discriminatory topic that blames the entire \textit{Muslim} community for the stone-pelting incident on the health workers. Topic 7 \& 8 are related to blaming the \textit{Tablighi Jamaat} event for the rise in COVID-19 cases and representing Islamophobic sentiments. An interesting observation comes from the number of tweets belonging to each of the topics - we found the hateful topics to be more representative of the data. This can be confirmed with the number of tweets belonging in each category - while topics in the non-hateful category have less than 10k tweets, almost 2-2.5 times more tweets belong to the last three hateful topics (6,7,8).

\begin{table*}[ht]%
\centering
\small
\begin{tabular}{|l|p{0.15\textwidth}|p{0.25\textwidth}||p{0.15\textwidth}|p{0.25\textwidth}|}
\hline
S.No. & \multicolumn{2}{|c||}{\textbf{Tabhlighi Jamaat Event (April 1st)}} & \multicolumn{2}{|c|}{\textbf{Occasion of Ramadan (April 25th)}} \\
\hline
\hline
& \multicolumn{1}{|c|}{\textbf{Topic}}& \textbf{Tokens}& \multicolumn{1}{|c|}{\textbf{Topic}}& \textbf{Tokens}\\
\hline
1 & India's efforts to contain virus foiled by Tablighis & every tablighi, indias efforts, severe setback, single cluster, efforts containing, keep counting  & Criticising media for spreading Islamophobia & hey islamophobics, islamophobics medicine, indian media, tablighijamaat, social media \\
\hline
2 & Islamic preacher praying to divert Covid-19 to Non-\textit{Muslim} nations & allah divert, nonmuslim nations, infections nonmuslim, divert covid19, preacher praying & Discriminatory behaviour towards \textit{Muslims} & muslims denied, pandemic differentiate, failure indian, denied food, racist rhetoric\\
\hline
3 & Islamophobia tainting India as doctors refuse to treat \textit{Muslim} patients & islamophobia taints, hospital refuses, muslim patient, crisis morality, news channels & Tablighi jamaat willing to donate plasma & plasma patients, members recovered, covid donate, cure others, tj members \\
\hline
\end{tabular}
\caption{ Some prevalent topics in \textit{Micro} topic modelling experiment for the entire corpus.}
\label{tab:micro_topic_modelling}
\end{table*}

For the \textit{Micro} topic modelling we chose two significant events -- 1) The release of reports related to the \textit{Tablighi Jamaat} event on 31st March/1st April 2020 ; 2) The occasion of \textit{Ramadan} on 25th April. The choice of these two events came from the fact that we wanted to look at the prevalent topics in two contrasting events on the stance towards the \textit{Muslim} community. Table~\ref{tab:micro_topic_modelling} presents the relevant topics for each of the event. For the \textit{Tablighi Jamaat} event, we see that most of the relevant topics are negatively portraying the \textit{Muslim} community and blaming them for the spread of COVID-19. Though there are topics (topic 3) where people are condemning Islamophobia and negligence in treating \textit{Muslim} patients after the Tablighi Jamaat incident, a majority of the tweets relate to how the Muslim community caused a setback to India's efforts to curb the virus.

On the contrary, topics associated with the occasion of Ramadan include many positive topics which talk about peace, humanity, and the good work that people have been doing during the pandemic, along with condemning the discriminatory behaviour against \textit{Muslims}.

Through this experiment, we analyzed the topics that were prevalent throughout the period of our study. While we observed topics in support of the discrimination against the \textit{Muslim} community, some others included condemnation of this discriminatory behaviour. Through this experiment, we were able to identify the different aspects of hate associated with \textit{Muslims} after going through a variety of topics. Additionally, the \textit{Micro} topic modelling helped us co-relate the prevalence of topics and events happening in the real world.

\section{User characteristics and communities}

\begin{figure*}[!htb]
\centering
\minipage{0.64\columnwidth}
\centering
  \includegraphics[width=0.75\columnwidth]{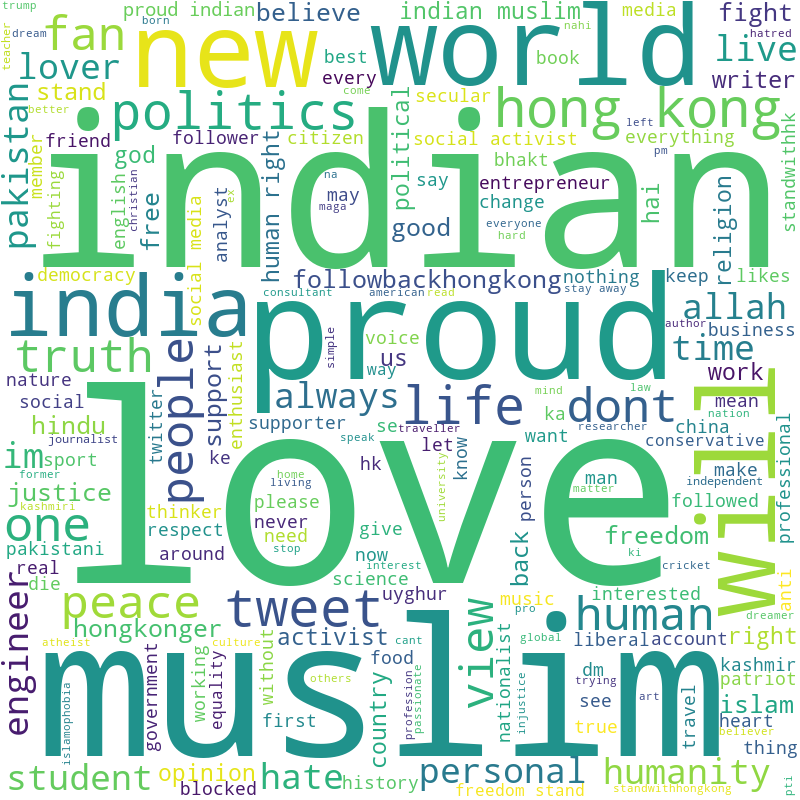}
  \caption{Word cloud generated from the bio of the users present in Class 1 ($0-25\%$ `Hateful' tweets). Added the color version to show you (will add the grayscale version in the main draft)}
 \label{fig:class1_bio}
\endminipage\hfill
\minipage{0.64\columnwidth}
\centering
  \includegraphics[width=0.75\columnwidth]{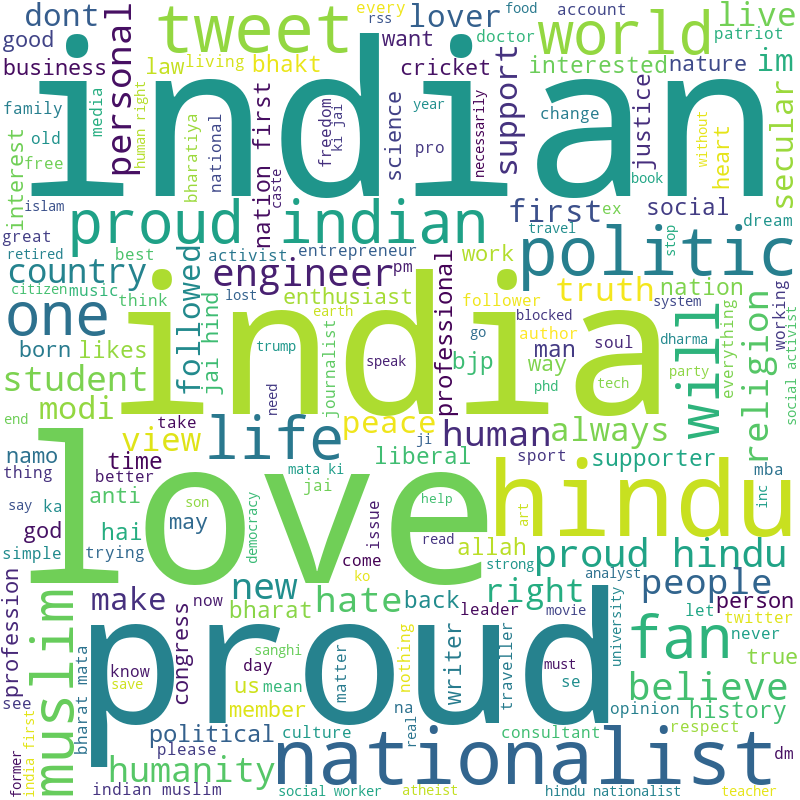}
  \caption{Word cloud generated from the bio of the users present in Class 2 ($25-50\%$ `Hateful' tweets). Added the color version to show you (will add the grayscale version in the main draft)}
  \label{fig:class2_bio}
\endminipage\hfill
\minipage{0.64\columnwidth}
\centering
 \includegraphics[width=0.75\columnwidth]{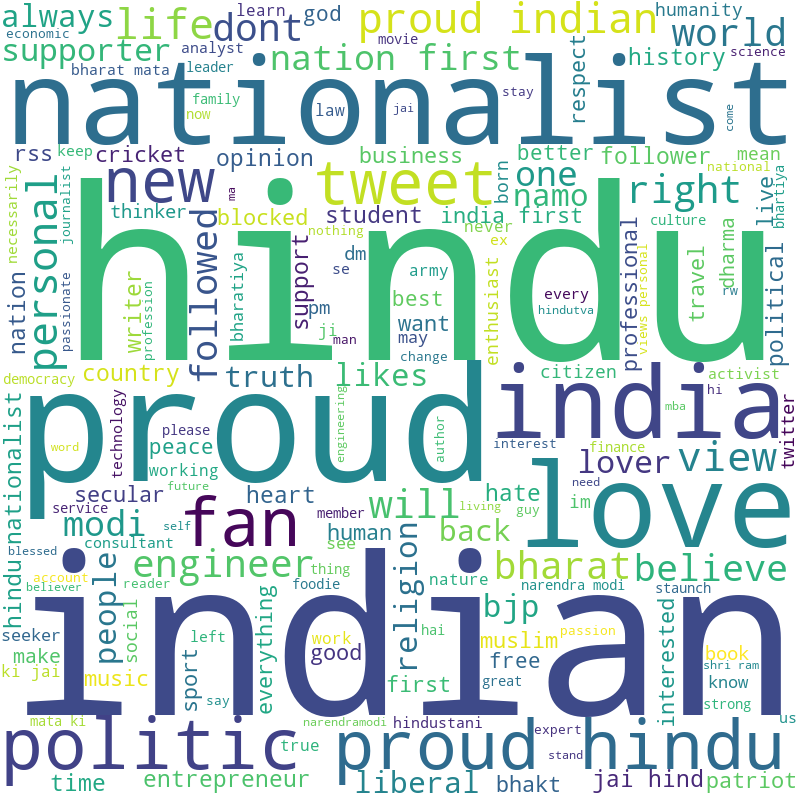}
  \caption{Word cloud generated from the bio of the users present in Class 3 ($50-100\%$ `Hateful' tweets). Added the color version to show you (will add the grayscale version in the main draft)}
 \label{fig:class3_bio}
\endminipage
\end{figure*}

The discourse on social media platforms is a reflection of how the users react to the offline events. Hence, it becomes essential to analyze the differentiating characteristics among the users. For this analysis, we categorized the users present in the \textit{CoronaBias} dataset in three categories based on the percentage of `Hateful' tweets among the total number of tweets tweeted by the user. To remove any instance of aberration, we only considered users who had tweeted at least five times during our study. Through this filtration process, we finally conducted this experiment on $12,328$ unique users. We categorized the users into the following classes: 1) \textbf{Class 1}: The percentage of hateful tweets is $ < 25\%$ of the total number of tweets by the user ($6,021$ users); 2) \textbf{Class 2}: The percentage of hateful tweets is $\geq 25\%$ and $ < 50\%$ of the total number of tweets by the user ($3,638$ users); 3) \textbf{Class 3}: The percentage of hateful tweets is $\geq 50\%$ of the total number of tweets by the user ($2,669$ users).

To answer \textbf{RQ3} relating to user characteristics and communities, we performed the following set of experiments:

\begin{itemize}
    \item We created word clouds from the user bios for each category of users. This experiment helped us analyze the ideologies for each set of users who actively engaged with the platform.

    \item We also created a user network graph to reveal the interactions between the users from different classes. Furthermore, we analyzed the pattern in the online activity of the users and co-related it to the offline events.
\end{itemize}

\subsection{Analysis of User Bio}

In this experiment, we wanted to analyze the differences in ideologies among the users and understand how they identified or presented themselves online. Hence, we extracted the user bios for the users in each category, pre-processed them, and generated word clouds for each class.

Figure~\ref{fig:class1_bio}, Figure~\ref{fig:class2_bio} and Figure~\ref{fig:class3_bio} presents the word clouds for Class 1, 2 and 3 respectively. As observed, when we move from Class 1 to Class 3, we see the growing significance of terms related to Nationalism and \textit{Hinduism}. A detailed look at the word cloud for Class 1 users revealed the usage of non-hateful/general terms like \textit{Proud Indian, Indian Muslim, Humanity, Muslim, Love} etc. This set of users had the least amount of hateful tweets against Muslims ( $ < 25\% $ ), and their user bios also indicated that many claimed to be Muslim and support ideologies of humanity and love. Out of the $6,021$, $4.9\%$ of user had the word \textit{Indian} written in their bio, $4.4\%$ of users used the word \textit{love} whereas $3.89\%$ of users used the word \textit{Muslim} in their bio.

Class 2 users' word cloud has many terms related to Nationalism/patriotism (\textit{India, India First, Proud Indian}) but at the same time also has religious terms (both \textit{Pro-Muslim and Pro-Hinduism}). $7.4\%$ of the $3,638$ users used the word \textit{Indian}, $6.9\%$ of users had the word \textit{Proud} whereas $5.2\%$ of users used the word \textit{Hindu} in their bio. As this is a transition class with medium level of hateful tweets against Muslims, the users' bios also indicated the same. In contrast to the prior two word clouds, Class 3 users' word cloud shows the sign of blatant \textit{Hinduism} which is portrayed as patriotism through the usage of Nationalist terminologies (\textit{Hindu Nationalist, Nation First, Proud Hindu}). The word cloud also contains terms related to Islamophobia (\textit{Chinese Jihadi, Ex Secular, Jihadi Products}). Moreover, we also observed a sharp rise in the percentage of users belonging to this class using these terms. $8.6\%$ of the $2,669$ users used \textit{Hindu}, $8.5\%$ used \textit{Proud} and $5.4\%$ used \textit{Nationalist} in their bio. 

This analysis revealed that religion was used as a symbol of patriotism by the far-right community present on Twitter. Moreover, the \textit{Muslim} community was portrayed as `anti-nationalist' due to the \textit{Tablighi Jamaat} incident. Overall, the user bio descriptions gave us insights into what ideologies users from each class believed in and we found a correlation between how they presented themselves and the nature of content they posted.

\subsection{User Network Graph and Activity Analysis}
As the first part of this experiment, we generated the follower-following network among the $12,328$ users. For creating the network graph, we considered each user as a node and added an edge between the nodes if one of the users (node A) followed the other user (node B). We removed isolated nodes i.e. nodes without any outgoing or incoming edges, to obtain $10,366$ nodes. We used the ForceAtlas2 layout algorithm \cite{forceatlas2014} on Gephi for the network spatialization with a scaling factor of 2.0 and gravity of 1.0. 

\begin{figure*}[!htb]
\centering
\minipage{0.80\columnwidth}
\centering
  \includegraphics[width=\columnwidth]{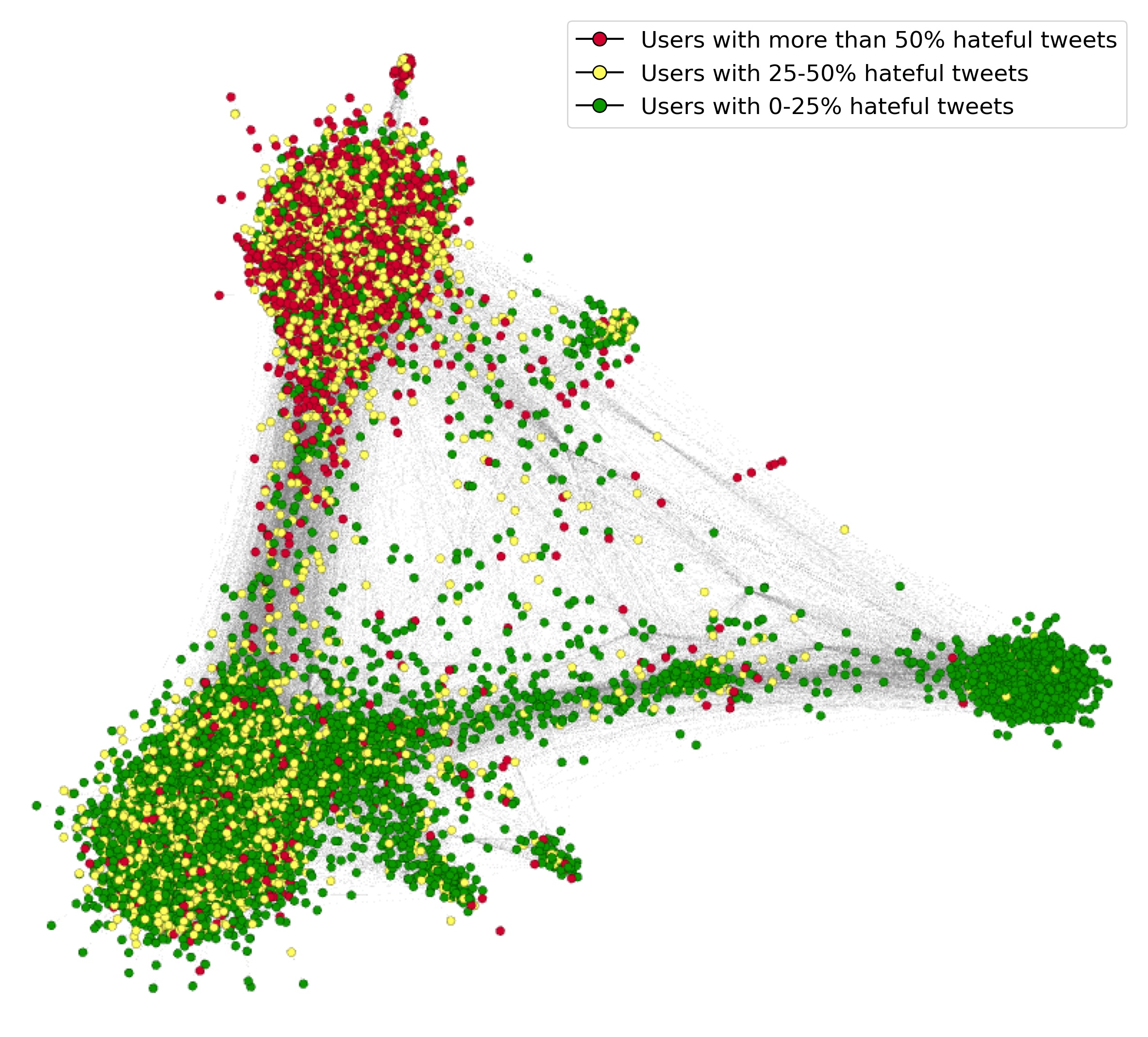}
  \caption{Network Graph for the case where the users are classified in one of three categories based on the percentage of `Hateful' tweets posted. Green nodes represent the users in Class 1 ($0-25\%$), Pink nodes represent the users in Class 2 ($25-50\%$) and Orange nodes represent the users in Class 3 ($50-100\%$). Added the color version to show you (will add the grayscale version in the main draft)}
  \label{fig:3class_network_graph}
\endminipage\hfill
\minipage{1.1\columnwidth}
\centering
  \includegraphics[width=\columnwidth]{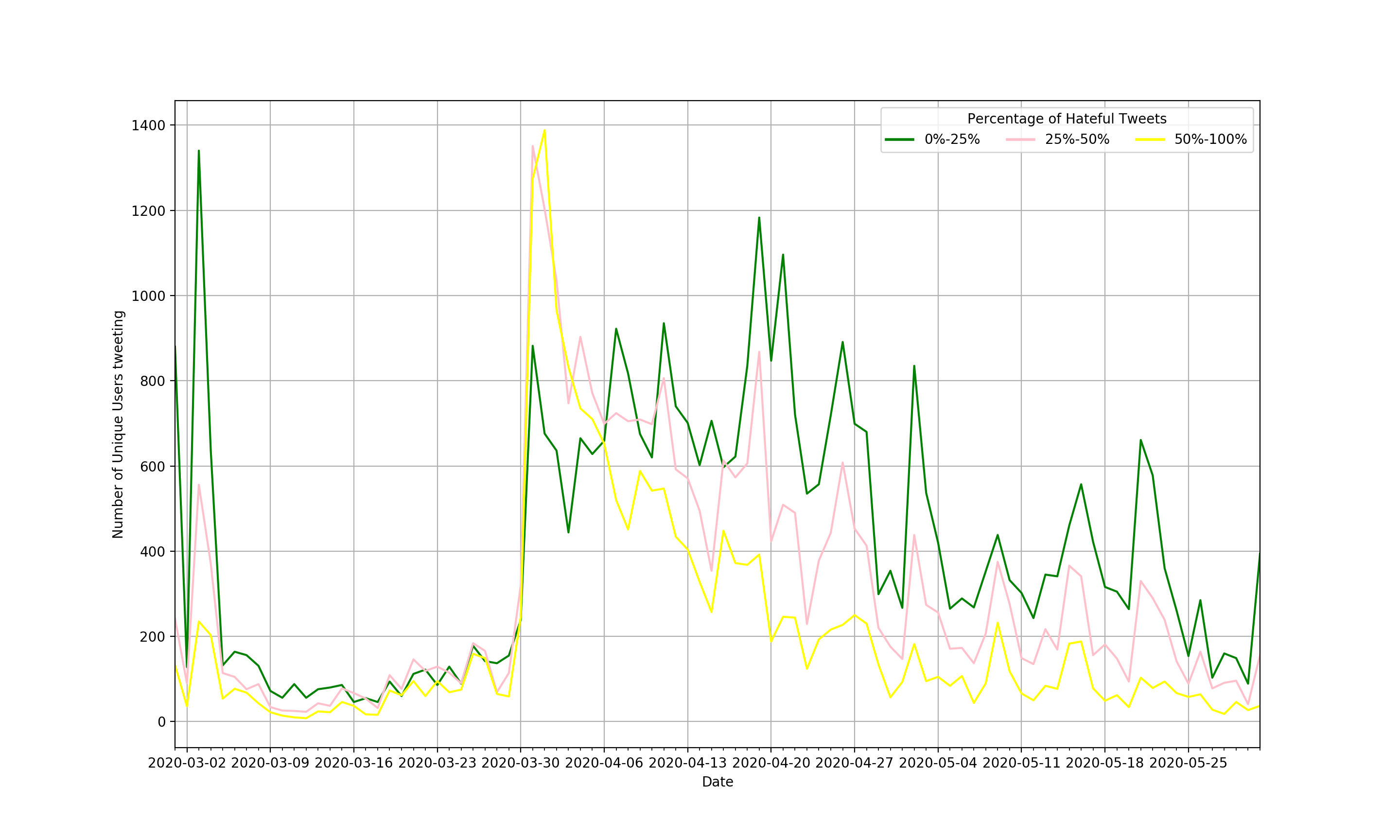}
  \caption{User Activity Graph for the case where the users are classified in one of three categories based on the percentage of `Hateful' tweets posted. Class 1 ($0-25\%$) in Green, Class 2 ($25-50\%$) in Pink and Class 3 ($50-100\%$) in Yellow.}
 \label{fig:class3_useractivity}
\endminipage
\end{figure*}


In Figure~\ref{fig:3class_network_graph} we found that users who posted a majority of hateful tweets (Class 3 coloured in red: users with $ \geq 50\% $ hateful tweets) are clustered closely together. The predominance of red colour is restricted to this one community implying that users that spread hate closely followed each other. This cluster does have some yellow nodes (Class 2: users with $ \geq 25\% $ and $< 50\%$ hateful tweets) but very few green nodes (Class 1: users with $ < 25\% $ hateful tweets), furthering our observation that this community of users spread more hateful content. We observed two other clusters of users that posted less hateful tweets. One cluster is predominantly green, and the other has majorly green nodes with some yellow nodes interspersed. This shows that there are also communities that posted majorly non-hateful content. To further validate the results obtained from our experiment, we ran Louvain community detection algorithm \cite{Blondel_2008} on our dataset. We obtained a total of $14$ separate communities based on the modularity class, out of which three communities constituted $99.7$\% of the graph. On further investigation we found these three communities to be congruous with the 3 major clusters we observed in Figure~\ref{fig:3class_network_graph}, where user demarcation was based on the percentage of hateful tweets posted.

Overall, we conclude that the follower-following relations between users are based on the kind of content they post and that the users in a cluster/community have a similar nature of the amount of hateful tweets, generating an echo-chamber effect overall. Some users of class 2 are scattered among the two major clusters, showing the equivocal behaviour of some users.

\begin{table*}[h]
    \centering
    \small
    \begin{tabular}{|p{0.1\columnwidth}|p{0.4\columnwidth}|p{0.2\columnwidth}|p{0.1\columnwidth}|p{0.4\columnwidth}|p{0.2\columnwidth}|}
    \hline
    S.No. & External URL Domain & Frequency & S.No. & External URL Domain & Frequency \\
    \hline
    1 & opindia.com & 9483 & 6 & indiatimes.com & 2865 \\
    \hline
    2 & aljazeera.com & 4032 & 7 & altnews.in & 2554 \\
    \hline
    3 & bbc.com & 3347 & 8 & thewire.in & 2250 \\
    \hline
    4 & jihadwatch.org & 3061 & 9 & dw.com & 2080 \\
    \hline
    5 & youtube.com & 2924 & 10 & washingtonpost.com & 2051 \\
    \hline
    \end{tabular}
    \caption{Top 10 most frequently referenced domains through URLs present in the tweets in the \textit{CoronaBias} dataset. The frequency denotes the total number of URLs that belong to these domains.}
    \label{tab:url_frequency}
\end{table*}

\begin{figure*}[ht]
\centering
\minipage{0.9\columnwidth}
  \includegraphics[width=\linewidth]{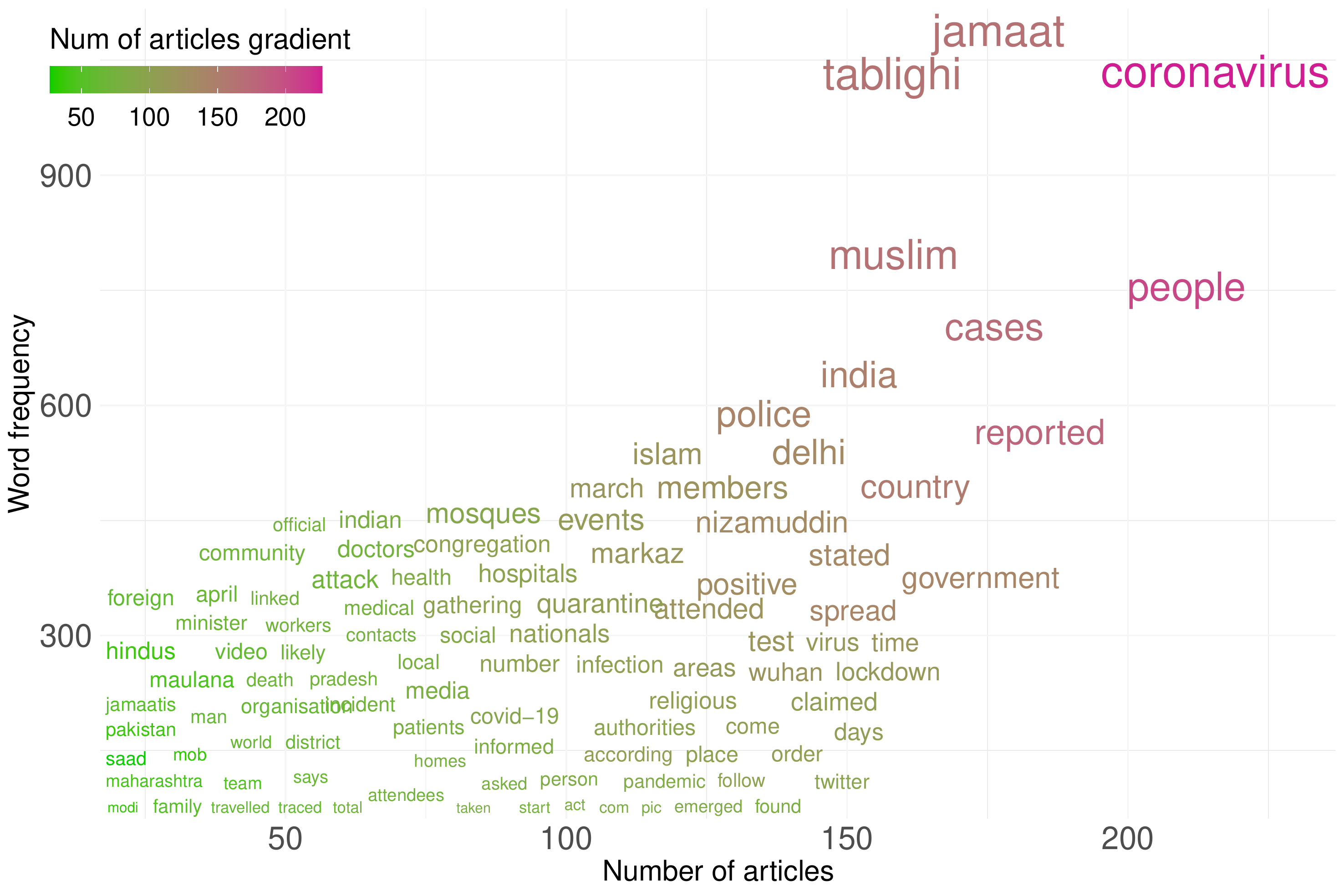}
  \caption{Chatter plot for article titles and content collected from OpIndia.}
  \label{fig:opindia_chatter}
\endminipage\hfill
\minipage{0.9\columnwidth}
  \includegraphics[width=\linewidth]{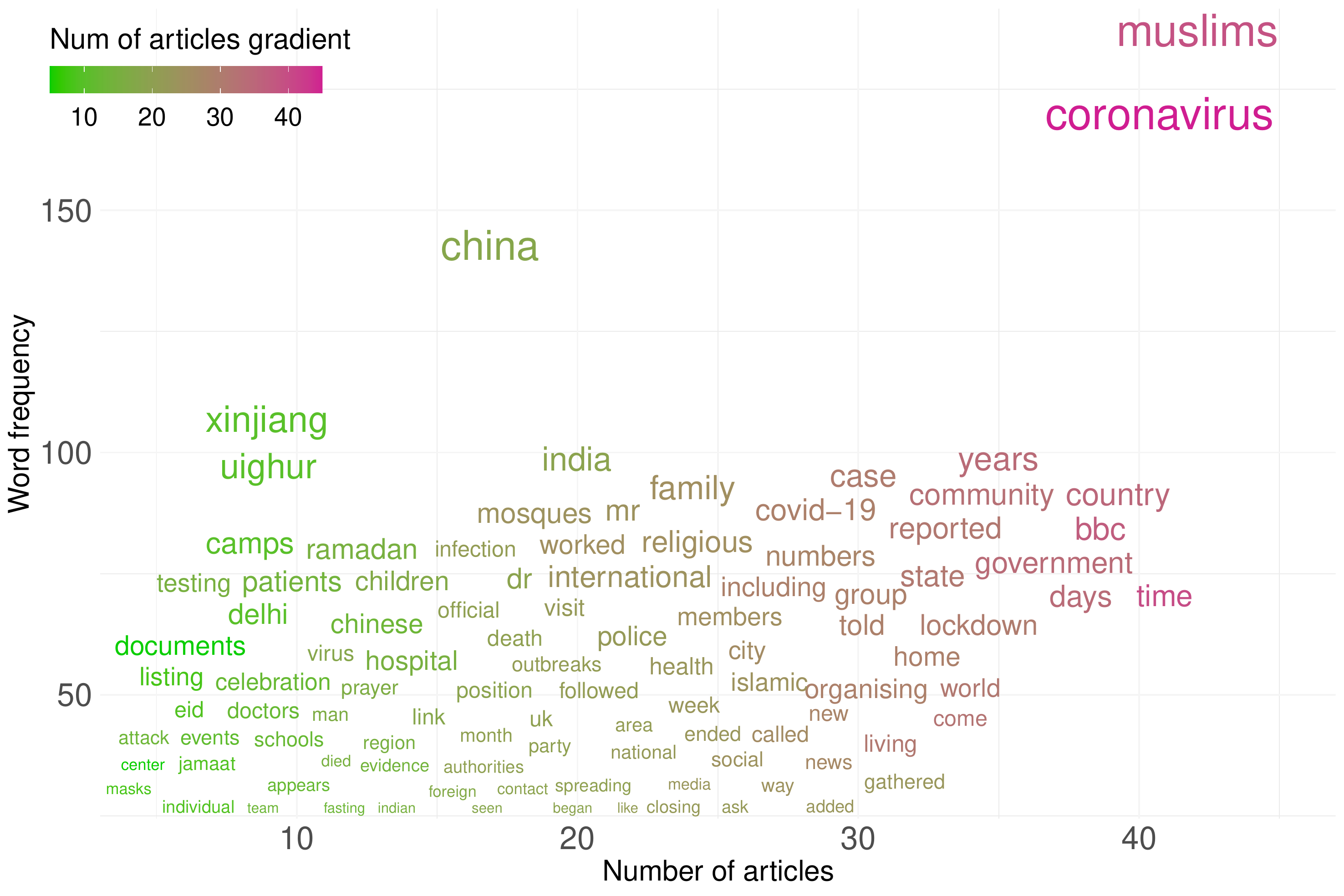}
  \caption{Chatter plot for article titles and content collected from BBC.}
  \label{fig:bbc_chatter}
\endminipage
\end{figure*}

\begin{figure*}[ht]
\centering
\minipage{0.9\columnwidth}
  \includegraphics[width=\linewidth]{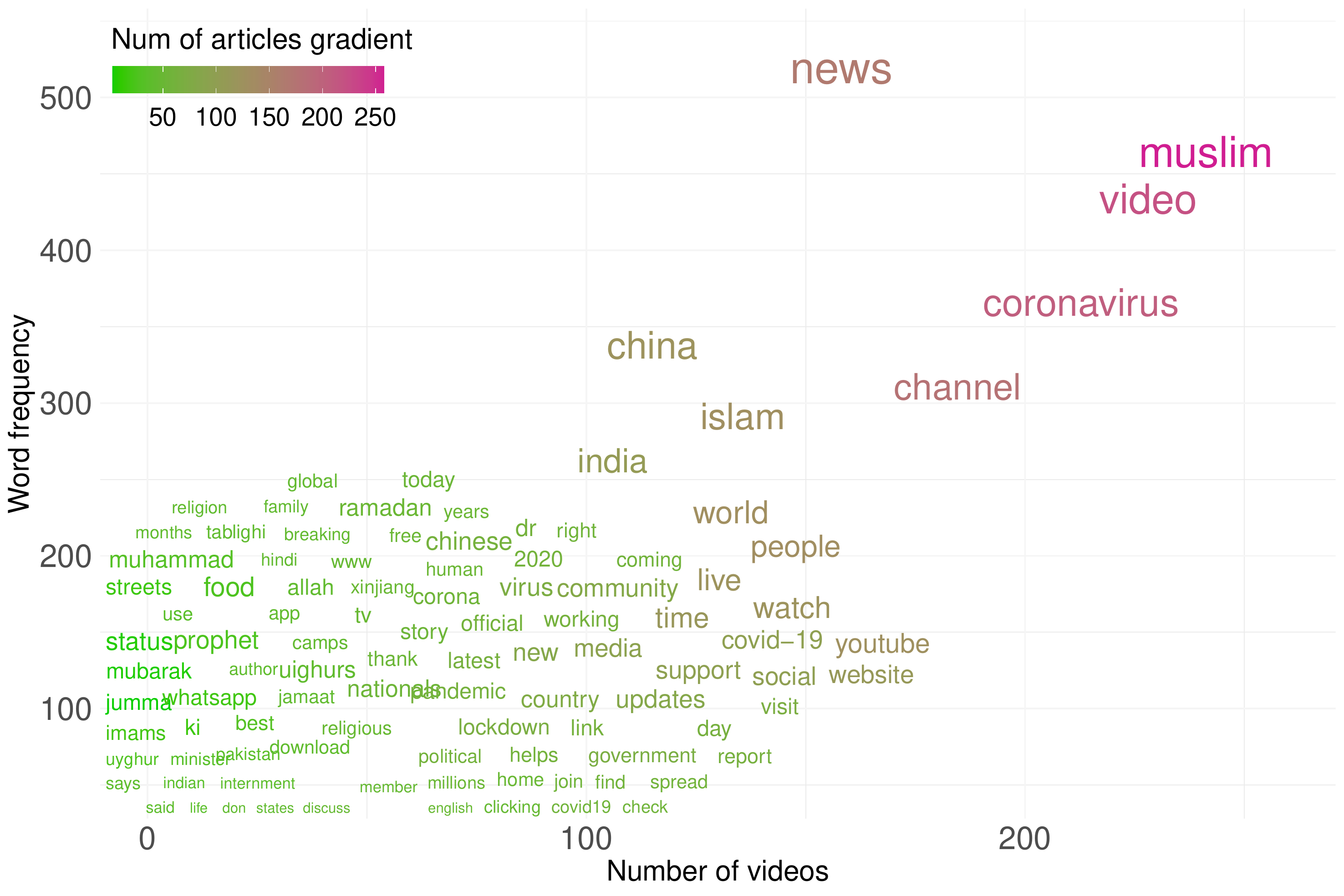}
  \caption{Chatter plot for the video titles and descriptions collected from Youtube.}
 \label{fig:Youtube_chatter_plot}
\endminipage\hfill
\minipage{0.9\columnwidth}
  \includegraphics[width=\linewidth]{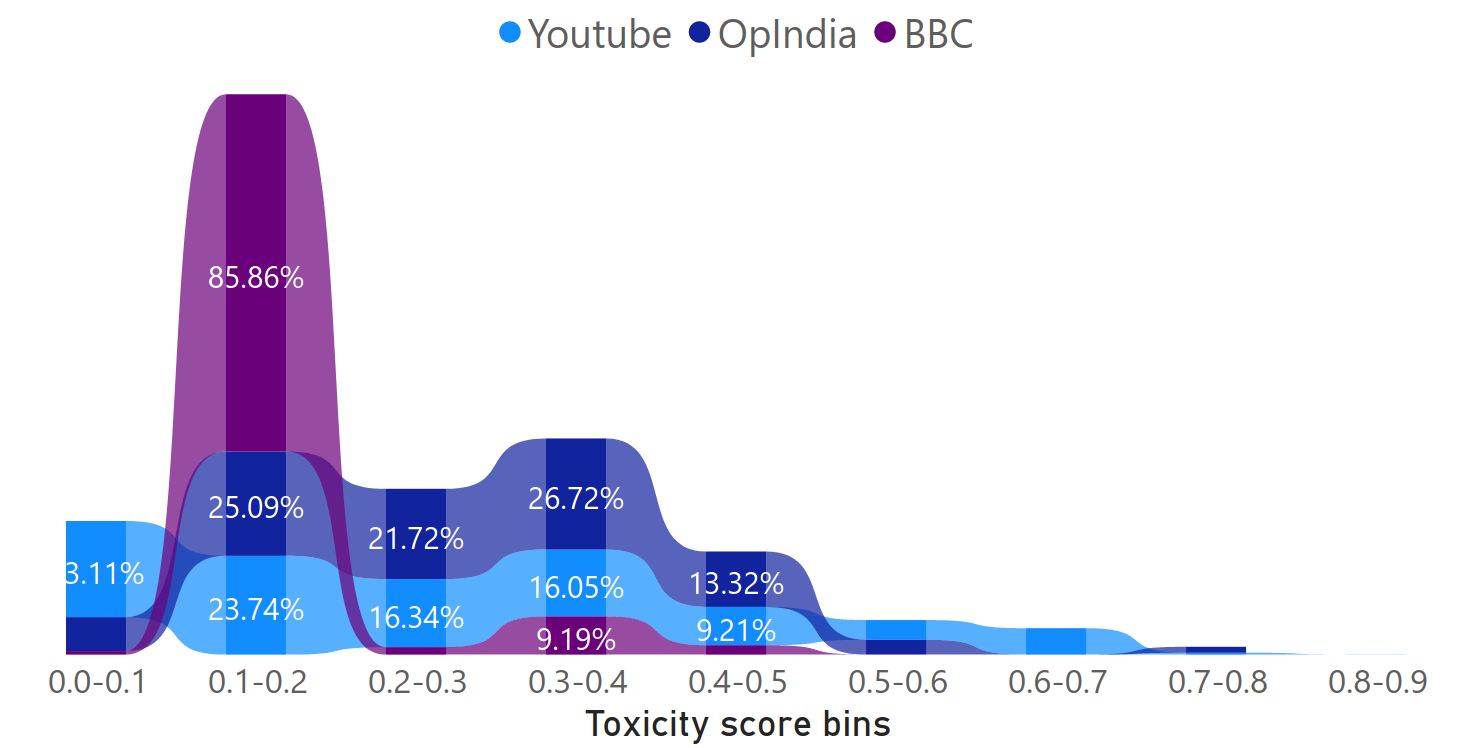}
  \caption{Ribbon plot for percentage of videos from Youtube, articles from OpIndia, and BBC in different Toxicity score bins obtained using Perspective API.}
  \label{fig:ribbon_plot}
\endminipage
\end{figure*}

Apart from the network graph, we also focused on observing the patterns in user activity during our study. We performed a temporal analysis of the number of unique users in each category posting tweets for each day to further reveal the differences in the behaviour. Figure~\ref{fig:class3_useractivity} presents the user activity graph for the three-class classification taxonomy based on \% of `Hateful' tweets. We observe a relatively high number of Class 1 unique users (users with $ < 25\% $ hateful tweets) active during the first week of March; we believe this was due to people posting \textit{Pro-Muslim} tweets on the issues of \textit{Uyghur Muslims} and rising COVID-19 cases in Iran. Furthermore, a subset of Class 1 users also formed a separate cluster in the network graph that focused on \textit{Muslims} in contexts other than the \textit{Tablighi Jamaat} event in India.
From the first week of March until 30th March, the user activity remained at normalcy.

31st March 2020 was when the reports about \textit{Tabhlighi Jamaat} congregation event were published. This day marked a sharp rise in the active users belonging to Class 2 and Class 3, suggesting the rise in Islamophobic content on Twitter. Interestingly, this incident also marked the start of behaviour where the users started posting more about the \textit{Muslim} community (both in a positive and negative context). We again observed a rise in activity for Class 1 users around the time of \textit{Ramadan} (last week of April). During this time, people started posting tweets with positive messages for the \textit{Muslim} community with hashtags like \textit{\#muslimsaviours}, lauding Muslims for their contribution to aiding with plasma therapy and distribution of food packages.

This experiment showed how particular events trigger a specific section of users to engage more with the platform. We also observed a pattern of increased discussion revolving around the \textit{Muslim} community during the ongoing pandemic.

\section{External URL Experiment}

Interestingly, the \textit{CoronaBias} dataset contains $\sim68.5\%$ of tweets with one or more URLs referencing external content. Hence, we focused on the analysis of URLs to answer \textbf{RQ4} relating to nature of external sources.

In this experiment, we extracted every URL from each of the unique tweets (excluding RTs). We then expanded each of the extracted URLs using the Requests\footnote{\url{https://requests.readthedocs.io/en/master/}} library and counted the frequency for each domain. Table~\ref{tab:url_frequency} presents the most popular external URL domains sorted by the number of tweets those links occur in. As expected, all but one domain (YouTube) belong to news media companies, as the users referenced these sources for the latest news headlines related to the disease spread, its prevention/cure, and associated lockdown policies. Among the popular social media platforms, we found the order as Facebook (579) followed by Instagram (322), Tumblr (155), and Linkedin (62). Surprisingly, we observed a large number of URLs from a few websites that have been known to disseminate anti-Muslim news ~\footnote{https://theprint.in/opinion/india-anti-muslim-fake-news-factories-anti-semitic-playbook/430332/} (OpIndia, Jihadwatch, and Swarajyamag). The frequency for URLs from OpIndia is maximum and double that of Aljazeera, which shows the scale of probable Islamophobic content shared through the URLs on Twitter.

In the second step, we digged deeper into the content referred through the URLs. For this experiment we chose three domains -- 1) OpIndia\footnote{\url{https://www.opindia.com/}}, 2) BBC\footnote{\url{https://www.bbc.com/}}, 3) Youtube\footnote{\url{https://www.youtube.com/}}. While we collected article title and article text data from OpIndia and BBC, we focused on title along with description for YouTube videos (ethical concerns related to data collection have been discussed in Section \ref{sec:ethical_concerns}). Fig~\ref{fig:opindia_chatter}, Fig~\ref{fig:bbc_chatter} and Fig~\ref{fig:Youtube_chatter_plot} present the chatter plots obtained for the data collected from OpIndia, BBC and YouTube respectively. As observed, the content referred from OpIndia talks about \textit{Muslims} and the \textit{Tablighi Jamaat} incident in great detail, with words like \textit{Tablighi, Jamaat} occurring more than 900 times across $\sim$ 180 articles. The content presented by BBC varied broadly while focusing on topics of Uighur \textit{Muslims}, general measures to prevent COVID-19, and lockdown due to the pandemic. The content referred from Youtube lay in between that of OpIndia and BBC.

Additionally, we analyzed the toxicity of external content. For this experiment, we used Perspective API \footnote{\url{https://www.perspectiveapi.com/}} to get the toxicity score for the data collected from the three sources. Perspective API defines toxicity as "rude, disrespectful, or unreasonable content". Figure~\ref{fig:ribbon_plot} presents the toxicity ribbon plot for each of the content from OpIndia, BBC, and Youtube. We divided the toxicity scores into bins 0-0.1, 0.1-0.2 etc, and plotted the percentage of articles from that domain belonging to a given toxicity bin. We observed that $\sim 86\%$ of the tweets containing a link from BBC lie in the bin 0.1-0.2 representing minor to no use of harsh words (here harsh words does not mean toxic words). In contrast to this, the number of articles from OpIndia were more evenly distributed in the range of 0.1-0.5. Moreover, the presence of a significant number of tweets in the range of 0.5-0.8 indicated the toxic nature of content present on OpIndia. Furthermore, occurrence of majority of articles from BBC in the 0.1-0.2 bin shows that any form of bias arising from the Perspective API due to religious terms like \textit{Muslims} etc. does not have a confounding effect on our claim of presence of toxic content for articles lying in the bin 0.5-0.8. For further validation of the results, we manually analyzed $50$ most frequently occurring articles referred from BBC and OpIndia in our dataset. We found that $\sim66\%$ of the $50$ most frequently occurring articles from OpIndia portrayed Islamophobic behaviour, whereas we did not find any Islamophobic article among the $50$ most frequently occurring articles from BBC.

The experiments performed in this section raise the concern that while the text/images presented in a tweet may not explicitly be hateful, tweets can still spread toxicity and hate from the URLs referenced in them. The widespread presence of media sources like OpIndia in our dataset, that frequently publish anti-Muslim content, shows that people used external sources to further Islamophobic views. The high percentage of tweets containing news media sources also highlights that people reference ongoing news in crucial times like a pandemic, where information becomes key.

\section{Conclusion and Discussion}

Through this work, we aim to provide a data-driven overlook of rising Islamophobia around the world. We curate a dataset of tweets related to the \textit{Muslim} community, known as ``\textit{CoronaBias}'' to study how anti-\textit{Muslim} discrimination starts and spreads during the pandemic caused by the Coronavirus. The dataset consists of 410,990 tweets from 224,229 unique users having an average of 34 words in each tweet. Using this dataset, we perform temporal analysis with the PELT experiment to draw a correlation between the offline events and the corresponding changes on Twitter. We find a growing association of the \textit{Muslim} community with the COVID-19 pandemic through the semantic similarity experiment. We perform \textit{Macro} and \textit{Micro} topic modelling to understand the topics prevalent during the course of our study in detail and during two focused windows. Qualitatively, we find a blend of mixed sentiments over the topics. Apart from the content-based analysis, we also perform a user-based experiment to reveal the differences in characteristics of various classes of users through their user bios and online activity. Finally, we conduct an external URL experiment to study the nature of the content referred to outside of Twitter.

Islamophobia has been studied broadly in the light of terrorism in the past. Researchers have shown that the Westerners often link Muslims to terror and violence \cite{ciftci2012islamophobia}, their policies result in an increase of Islamophobia \cite{alam2013islamophobia}. In addition,
\cite{briskman2015creeping} discuss that media, government, and community discourses converge to promote Islam as dangerous and deviant. Through this work, we create an understanding of anti-Muslim sentiments that are not directly coherent with terrorism but can harm the community in a dire manner. The spread of negative sentiments towards any community is unacceptable; therefore, we wish to build on this work to explore its mitigation. 





\bibliographystyle{ACM-Reference-Format}
\bibliography{sample-base}


\end{document}